\documentclass[11pt]{article}
\usepackage{amsmath,epsf,amssymb,latexsym,enumerate,shadow,array,color,bbold}
\usepackage[nosort]{cite}
\usepackage{slashed}
\usepackage{graphicx,wasysym}

\DeclareFontFamily{U}{rsf}{} \DeclareFontShape{U}{rsf}{m}{n}{
  <5> <6> rsfs5 <7> <8> <9> rsfs7 <10-> rsfs10}{}
\DeclareMathAlphabet\Scr{U}{rsf}{m}{n} \makeatletter
\@addtoreset{equation}{section} \makeatother

\def\be{\begin{equation}}
\def\ee{\end{equation}}
\def\ba{\begin{array}}
\def\ea{\end{array}}

\newcommand{\bea}{\begin{eqnarray}}
\newcommand{\eea}{\end{eqnarray}}

\def\K{K\"{a}hler}
\newcommand{\hc}{{\rm h.c.}}
\newcommand{\ft}[2]{{\textstyle\frac{#1}{#2}}}
\def\rmi{{\rm i}}
\def\rmd{{\rm d}}

\newsavebox{\uuunit}
\sbox{\uuunit}
    {\setlength{\unitlength}{0.825em}
     \begin{picture}(0.6,0.7)
        \thinlines
        \put(0,0){\line(1,0){0.5}}
        \put(0.15,0){\line(0,1){0.7}}
        \put(0.35,0){\line(0,1){0.8}}
       \multiput(0.3,0.8)(-0.04,-0.02){12}{\rule{0.5pt}{0.5pt}}
     \end {picture}}
\newcommand {\unity}{\mathord{\!\usebox{\uuunit}}}
\usepackage{ifpdf}
\ifx\pdfoutput\undefined
   \pdffalse
\else
   \pdfoutput=1
   \pdftrue
  \usepackage[pdftex]{hyperref}
\def\u0{{\underline 0}}

\def\url{{\underline {r+\ell}}}

\newcommand{\rf}[1]{(\ref{#1})}

\def\rmi{{\rm i}}
\def\rmd{{\rm d}}

\def\ib{{\bar \imath}}
\def\jb{{\bar \jmath}}
\newcommand{\tGamma}{{\tilde\Gamma}}

\newcommand{\zp}{0}
\newcommand{\zm}{{\bar 0}}
\newcommand{\ip}{i}
\newcommand{\jp}{j}
\newcommand{\im}{\ib}
\newcommand{\jm}{\jb}
\newcommand{\pone}{1}
\newcommand{\ptwo}{2}
\newcommand{\pthree}{3}
\newcommand{\mone}{\bar 1}
\newcommand{\mtwo}{\bar 2}
\newcommand{\mthree}{\bar 3}

\newcommand{\lp}{\left(}
\newcommand{\ls}{\left[}
\newcommand{\rp}{\right)}
\newcommand{\rs}{\right]}

\begin{document}

\begin{titlepage}

\begin{flushright}
TUW-15-04\\
RUG-15-16
\end{flushright}

\hskip 0.5cm

\vskip 0.5cm

\begin{center}

{\huge {$\overline {\rm \bf D3}$ {\bf and dS}}}

\

\

 {\Large \bf Eric A. Bergshoeff$^1$},  {\Large \bf Keshav Dasgupta$^{2,3}$},   {\Large \bf Renata Kallosh$^3$},\\

\vskip 0.5 cm

   {\Large \bf Antoine Van Proeyen$^4$}\   {\Large \bf and  Timm Wrase$^5$} \vskip 0.5cm
{\small\sl\noindent
$^1$ Van Swinderen Institute for Particle Physics and Gravity,\\
University of Groningen, Nijenborgh 4, 9747 AG Groningen, The Netherlands\\\smallskip
$^2$ Ernest Rutherford Physics Building, McGill University,\\
3600 University Street, Montr\'{e}al QC, Canada H3A 2T8\\\smallskip
$^3$ Department of Physics, Stanford University, Stanford, CA
94305 USA \\\smallskip
$^4$ Instituut voor Theoretische Fysica, KU Leuven,\\ Celestijnenlaan 200D, B-3001 Leuven,
Belgium \\\smallskip
$^5$ Institute for Theoretical Physics, TU Wien
Wiedner Hauptstr. 8-10, A-1040 Vienna, Austria}

\end{center}

Email: \texttt{e.a.bergshoeff@rug.nl}, \texttt{keshav@hep.physics.mcgill.ca}, \texttt{kallosh@stanford.edu},

\vspace{-.3cm} \hspace{1.1cm} \texttt{Antoine.VanProeyen@fys.kuleuven.be}, \texttt{timm.wrase@tuwien.ac.at}

\vskip 1 cm

\begin{abstract}

\

The role of the $\overline {\rm D3}$ brane in providing de Sitter vacua with spontaneously broken supersymmetry in the KKLT construction is clarified. The first step in this direction was explained in \cite{Kachru:2003aw,Kachru:2003sx}: it was shown there that in the GKP background the bosonic contributions to the vacuum energy from the DBI and WZ term cancel for a D3 brane, but double for a  $\overline {\rm D3}$ brane, leading to de Sitter vacua. The next step was taken in \cite{Kallosh:2014wsa} where the analogous mechanism of the doubling (cancelation) of the $\overline {\rm  D3}$  (D3) DBI and WZ terms was discovered  in the presence of Volkov-Akulov fermions living on the brane, in a flat supergravity background. Here we confirm this mechanism of doubling/cancelation for the $\overline {\rm  D3}$/D3 brane in the GKP supergravity background preserving $\mathcal{N}=1$, $d=4$ supersymmetry. We find that imaginary self-dual $G_{(3)}$ flux of type $(2,1)$ nicely removes the $SU(3)$ fermion triplet by giving it a large mass, while leaving the Volkov-Akulov goldstino, which is the $SU(3)$ singlet, massless. This makes the de Sitter landscape in D-brane physics clearly related to de Sitter vacua in effective $d=4$ supergravity with a nilpotent multiplet and spontaneously broken  supersymmetry.

\end{abstract}

\vspace{24pt}
\end{titlepage}

%\tableofcontents

\newpage

\section{Introduction}

The role of the $\overline {\rm D3}$ brane in the presence of an O3-plane in providing an uplift from supersymmetric AdS vacua to dS vacua in the KKLT construction \cite{Kachru:2003aw} was explained in detail in \cite{Kachru:2003sx}. The bosonic D3 and $\overline {\rm  D3}$ actions were studied there in the curved background $AdS_5\times X_5$, with $X_5$ an Einstein manifold. The action at a fixed position of the brane or anti-brane in the extra dimensions, $r_1(\sigma)=r_0$, has the following  form \cite{Kachru:2003sx}
\be
S^q=S_{DBI}+ q \, S_{WZ} = - T_3 \int  d^4\sigma\, \sqrt {-g}  \Big ({r_0\over R}\Big)^4   + q \,T_3 \int \, C_4 \, .
\label{bos}\ee
Here $q=1$ for a D3 brane and $q=-1$ for a $\overline {\rm  D3}$ brane, and $R$ is  the characteristic length scale of the $AdS_5$ geometry. The 4-form in this case is  $C_4 =\Big ({r_0\over R}\Big)^4 d^4 \sigma$. It was  observed in \cite{Kachru:2003sx} that at any fixed position
\be
S^{\rm D3}=0\, ,  \qquad S^{\overline {\rm  D3}}= - 2T_3  \Big ({r_0\over R}\Big)^4\int d^4 \sigma \sqrt {-g}  \,.
\label{CancDoubl}\ee
This leads to an effective positive energy for the $\overline {\rm  D3}$ at position $r_0$ in a background with unbroken supersymmetry, so that
\be
V= 2T_3    \Big ({r_0\over R}\Big)^4\,,
\label{pot1} \ee
which can uplift the vacuum to a dS one. The same feature takes place in the GKP background \cite{Giddings:2001yu}. The metric in such a background is of the form
\begin{equation}
  ds_{10}^2 = e^{2 A(z)} \eta_{\mu\nu} dx^\mu dx^\nu + e^{-2A(z)} \tilde{g}_{i \jb} dz^i d\bar{z}^{\jb}\,, \qquad \mu,\nu=0,1,2,3, \qquad i, \jb = 1,2,3,
 \label{metricGKP}
\end{equation}
and the self-dual 5-form flux is given by
\begin{equation}
  \tilde{F}_5 =(1+*_{10})  [ d\alpha(z) \wedge dx^0 \wedge dx^1 \wedge dx^2 \wedge dx^3]\,.
 \label{F5GKP}
\end{equation}
The equations of motion require that $e^{4 A(z)}= \alpha(z)$, so that the pull-back of $C_4$ is given by
$C_4= \alpha(z)\ d^4 \sigma= e^{4A(z)} \ d^4 \sigma$. When the $\overline {\rm  D3}$ brane is located at some $z=z_0$ we have
\be
V= 2T_3  e^{4A_0}\,,\qquad \mbox{with}\qquad A_0 \equiv A(z_0) \,.
\label{pot2}
\ee
One may view this observation as a first indication that the complete brane action with account of fermions which live on the brane, might exhibit an analogous phenomenon: cancelation of the DBI and WZ terms for the D3 brane and doubling for the $\overline {\rm  D3}$ brane in the GKP background \cite{Giddings:2001yu} preserving $\mathcal{N}=1$ supersymmetry.

The analysis in \cite{Kachru:2003aw,Kachru:2003sx} is based on the bosonic action of the brane.
Meanwhile, one can start instead with the full $\kappa$-symmetric actions of the D3 and $\overline {\rm  D3}$ brane, which include the fermions living on the (anti-) brane \cite{Cederwall:1996pv,Bergshoeff:1996tu,Aganagic:1996nn,Bergshoeff:1997kr,Kallosh:1997aw, Simon:2011rw,Bergshoeff:2013pia}.
These actions after gauge-fixing of the local fermionic $\kappa$-symmetry have spontaneously broken supersymmetry in the presence of the fermions remaining on the brane after the gauge-fixing.
The corresponding analysis of the actions was performed in \cite{Kallosh:2014wsa} in a flat supergravity background \footnote{The mechanism of doubling/cancelation of the $\overline {\rm  D9}$/D9 branes was first discovered and studied in \cite{Bergshoeff:1999bx} in the flat superspace background and in \cite{Riccioni:2003ga} in the curved superspace background, at the level quadratic in fermions. Our results for the $\overline {\rm  D3}$/D3 branes are in agreement with the ones following from a compactification and performing T-dualities of the $\overline {\rm  D9}$/D9 branes.}. It was shown that in the presence of fermions, under certain orientifolding conditions, the cancelation of the DBI and WZ terms on the D3 brane and doubling on the $\overline {\rm  D3}$ take place in the exact non-linear expression with fermions. This suggests the existence of a supersymmetric version of the KKLT uplifting. The fermion action is a Volkov-Akulov goldstino action \cite{Volkov:1973ix}, as  expected  according to \cite{Kallosh:1997aw}, in a flat background.
The result of \cite{Kallosh:2014wsa} is
\be
S^{\rm D3}=0\, ,  \qquad S^{\overline {\rm  D3}}= - 2T_3 \int d^4 \sigma \det E  = - 2T_3 \int  E^0\wedge E^1 \wedge E^2  \wedge E^3\,,
\label{eq:DoublingFerm}
\ee
where $E^a = \delta^a_m dx^m + \bar \lambda \gamma^a d\lambda$ for $a=0,1,2,3$. In this form of the action the linearly realized supersymmetry is manifest since $E^a $ is invariant under simultaneous transformations of $x^m(\sigma)$ and $\lambda(\sigma)$.  However, the $\mathcal{N}=4$ supersymmetry of this action is broken spontaneously in the gauge in which the reparametrization symmetry on the brane is fixed and $x^m=\sigma^m$. The non-linearly realized supersymmetry acts only on the fermion
\be\label{eq:lambda}
\delta_\zeta \lambda =  \zeta +  \bar \lambda \gamma^a \zeta \partial_a \lambda\,,
\ee
and the action is invariant.
In the absence of the fermion $\lambda$ the above action reproduces the positive potential \rf{pot2} with $e^{4A_0}=1$ since in a flat background \cite{Kallosh:1997aw} with $e^a_m =\delta^a_m$ we have $e^{A_0}= 1$.

For simplicity, in \cite{Kallosh:2014wsa}, following \cite{Aganagic:1996nn}, we have restricted our analysis to the case of a flat supergravity background without fluxes, in the hope that it then will be possible to generalize the analysis to the case of a curved supergravity background with fluxes. The purpose of this paper is exactly this: to study the actions of the D3 and $\overline {\rm  D3}$ brane in a curved space GKP background with unbroken $\mathcal{N}=1$ supersymmetry. We would like to find out if the doubling/cancelation that was found in \cite{Kachru:2003sx} (cf. \eqref{CancDoubl}) and in \cite{Kallosh:2014wsa} (cf. \eqref{eq:DoublingFerm}) persists in the supersymmetric brane actions in a GKP background. The answer is positive and  provides a convincing picture of the KKLT uplifting to de Sitter vacua with spontaneously broken supersymmetry.

The KKLT construction of de Sitter vacua in string theory proposed in 2003, was a quick response to the discovery of the dark energy. Now, a  decade later, the experimental confirmation  of the dark energy is  much stronger than before, it is a cornerstone of the cosmological concordance $\Lambda$CDM model with 69\% of dark energy. According to the 2015 release of Planck \cite{Planck:2015xua} the equation of state of dark energy is now constrained to $w=- 1.006 \pm 0.045$, combining Planck data with other astrophysical data, including Type Ia supernovae.
 This is  consistent with the expected value for a cosmological constant, i. e. with a de Sitter vacuum with $w=-1$. Comparing with the  2013 release  of the Planck data \cite{Ade:2013zuv} where  the dark energy equation of state parameter was constrained to be $w=- 1.13 \pm 0.13$, we see a significant improvement during the last two years.
 Therefore, the  importance of the correct understanding of the role of the $\overline {\rm D3}$ brane dS uplifting is increasing:
one would like to describe a situation where the physics of D-branes can help us to explain the dark energy of the universe which we observe.

A new analysis of the stability issues of the $\overline {\rm D3}$ brane dS uplifting was recently performed in \cite{Michel:2014lva,Hartnett:2015oda,Danielsson:2015eqa}. In \cite{Michel:2014lva} the focus is on the simplest case of a single anti-brane and the authors use effective field theory to gain a better understanding of the anti-brane dynamics in flux backgrounds. The conclusion is reached  that this analysis gives a valid description of the anti-brane, and that there is no instability in this approximation. In \cite{Hartnett:2015oda}  new studies were performed of the situation when the brane polarization smoothens out the flux singularity, preventing brane-flux decay.  A suggestion of a new potential decay channel was proposed in \cite{Danielsson:2015eqa}, however, an explicit consistent construction is not  available.

Our approach is complementary to recent studies in \cite{Michel:2014lva,Hartnett:2015oda}: we study the $\overline {\rm D3}$ brane action with the explicit fermions on the world-volume. This leads to a description in terms of spontaneously broken $\mathcal{N}=1$ effective $d=4$ supergravity which has a novel feature: a nilpotent goldstino superfield that gives a supersymmetric description of the $\overline {\rm D3}$ brane uplifting term. We also find it useful in our analysis to place the $\overline {\rm D3}$ brane on top of an O3 plane that removes the worldvolume scalars, and thereby any potential tachyons.

The paper is organized as follows.  Sec.~\ref{ss:D3aD3curved} provides background material for presenting our new results in Sec.~\ref{ss:bD3D3superspace}. Namely, we provide a short review of the Volkov-Akulov (VA) supersymmetry \cite{Volkov:1973ix}, which is a central point in the relation between D-branes and effective supergravity models that have natural de Sitter vacua with spontaneously broken supersymmetry. The corresponding supermultiplet involves only a fermionic field, the goldstino,  and has no bosons.
We continue, following \cite{Bergshoeff:1996tu}, with a review of the $\kappa$-invariant worldvolume actions in a curved superspace background and the role of  DBI and WZ parts of the action: they  are separately supersymmetric but cancel each other contributions to local $\kappa$-symmetry variations. We continue with a short review of the mechanism of the cancelation of the fermion-dependent terms in the DBI and WZ part of the D3 brane and the doubling for the $\overline {\rm D3}$ brane in a flat superspace background  when the orientifolding condition is imposed. This leaves us with a VA action for the 16-component goldstino of the gauge-fixed $\overline {\rm D3}$ brane action. This 16-component 10d spinor can be decomposed into four 4d Dirac spinors,  a singlet and a triplet under the $SU(3)$ holonomy group of the internal Calabi-Yau manifold.

In Sec.~\ref{ss:bD3D3superspace}  we  introduce a curved superspace background via the supervielbein and fluxes and provide strong evidence that the effect of cancelation/doubling discovered in the flat background in \cite{Kallosh:2014wsa} is preserved.\footnote{Namely, the curved superspace versus the flat one has two features, one is purely geometrical and involves a curved supervielbein, the other is the appearance of the form-fields. The account of a supervielbein versus a flat one is discussed in the beginning of Sec.~\ref{ss:bD3D3superspace} where we explain why the corresponding generalization of the flat DBI and WZ term are such that to all orders in fermions they either cancel or combine. This is due to the uniqueness of the $\det E$ expression. The other feature we have to take into account are the form-fields. These we analyze only at the level quadratic in fermions.} In Sec.~\ref{sec:quadratic} we focus our attention on terms quadratic in fermions in the gauge-fixed $\overline {\rm D3}$ brane action in a bosonic  GKP background \cite{Giddings:2001yu} preserving $\mathcal{N}=1$ supersymmetry. We find that the triplet of fermions acquires a mass due to the presence of the ISD flux whereas the singlet 4d goldstino remains massless. This accomplishes the derivation of the VA goldstino action with spontaneously broken $\mathcal{N}=1$ supersymmetry. It would be interesting to extend the analysis of the action with account of fluxes to the level beyond quadratic in fermions. Since we allow only the primitive (2,1) flux  which carries internal indices, it will likely be contracted with the triplet spinors at higher orders. But since they are heavy on the basis of quadratic approximation, we will cut them off anyway. Thus, the full-fermion analysis will likely confirm our expectation that the cancelation/doubling in the GKP background is valid to all orders in singlet fermions, with account of the allowed fluxes.

In the discussion Sec.~\ref{ss:discussion} we  relate   the $\overline {\rm  D3}$ brane action in the GKP background, preserving $\mathcal{N} =1$ supersymmetry to $\mathcal{N}=1$ supergravity with de Sitter vacua and spontaneously broken $\mathcal{N}=1$ supersymmetry of the VA type. In Appendix~\ref{app:d10d4} we provide the details of the spinor reduction from 10d to 4d.  In Appendix~\ref{app:example} we give an explicit example of the $\overline {\rm D3}$ brane action in a simple toroidal GKP background. Finally, in Appendix~\ref{ss:nogo} we revisit the no-go theorem by Gibbons-Maldacena-N\'{u}\~{n}ez \cite{GMN} and discuss how consistent de Sitter vacua are realized in our framework.

\section{The D3 and \texorpdfstring{$\overline {\rm \bf D3}$}{Bar D3} brane actions in curved superspace}
  \label{ss:D3aD3curved}

 \subsection{A short review of Volkov-Akulov supersymmetry}

The interest in the Volkov-Akulov supersymmetry \cite{Volkov:1973ix} was renewed recently because of its application to cosmology which started in \cite{Antoniadis:2014oya}. In \cite{Ferrara:2014kva} it was  shown that for the general case the corresponding chiral nilpotent multiplet studied earlier in \cite{rocek,Komargodski:2009rz} can be defined in the superconformal version of supergravity. Interesting applications to cosmology where inflationary models have a spontaneously broken supersymmetry of the VA type at the minimum of the potential were developed in  \cite{Kallosh:2014via}.

There is a generic expectation in $d=4$, based on linearly realized $\mathcal{N}=1$ supersymmetry of the Golfand-Likhtman-Wess-Zumino type,  that every boson has a fermion as partner, and vice versa. However, the non-linearly realized supersymmetry of the Volkov-Akulov type which presents a spontaneously broken $\mathcal{N}=1$ supersymmetry, involves only a self-interacting fermion, the goldstino, which has no bosonic partner. The meaning of  a spontaneously broken $\mathcal{N}=1$ supersymmetry in this case is the following: there exist an action, depending on one 4-component spinor, which is invariant under a supersymmetry transformation. The transformation contains a constant spinor and a term quadratic in the fermionic field. The invariant action can be presented in the form \rf{eq:DoublingFerm} with fixed reparametrization
symmetry, $x^m=\delta^m_\mu \sigma^\mu$,
\be
S_{VA}= - M^4  \int d^4 \sigma \det E  = - M^4  \int E^0\wedge E^1 \wedge E^2  \wedge E^3
\,, \qquad  E^a = \delta^a_\mu d\sigma^\mu + \bar \lambda \gamma^a d\lambda\,,
\label{VA}\ee
where $M$ is a mass parameter and $M^4$ is not necessarily equal to the $2 T_3$ in equation \eqref{eq:DoublingFerm}.
The action depends only on the fermion $\lambda(\sigma)$ and is invariant under the  $\mathcal{N}=1$ supersymmetry transformation given in \eqref{eq:lambda}. It can also be presented in a form with manifestly realized linear supersymmetry, as shown in \cite{Ferrara:2014kva} in eq. (2.20)
\begin{equation}
{\cal L}_{VA} \ = \ \bigg[ S\, \overline{S} \bigg]_D \ + \ \bigg[M^2 S \,+\Lambda S^2
\bigg]_F \,, \label{va2}
\end{equation}
where $\Lambda$ is a Lagrange multiplier chiral superfield. After the equations of motion for the Lagrange multiplier are solved, with $S^2(x, \theta)=0$, one finds that the action takes the form given in \cite{Komargodski:2009rz}
\be
{\cal L}_{VA}= - M^4 +\rmi \partial_a \bar \psi \bar \sigma^a \psi + {1\over 4 M^4} \bar \psi^2 \partial^2 \psi^2 - {1\over 16 M^{12}} \psi^2 \bar \psi^2 \partial^2 \psi^2 \partial^2 \bar \psi^2 \,,
\label{VA1}\ee
with $\bar{\sigma}^a = (-\unity, -\sigma_n)$. As it is shown explicitly in \cite{Kuzenko:2010ef}, the above action agrees with the original VA action \eqref{VA} after a spinorial field redefinition  $\psi=M^2 \lambda$ plus terms non-linear in fermions.

 \subsection{\texorpdfstring{$\kappa$}{Kappa}-invariant worldvolume actions}

The worldvolume actions for D3 brane solutions of supersymmetric field theories may be viewed as 4-dimensional non-linear sigma-models with a superspace as the target space \cite{Hughes:1986fa}. In the notation of \cite{Bergshoeff:1996tu} the worldvolume fields $Z^M (\sigma)$ define a map from the worldvolume with coordinates $\sigma^\mu$ ($\mu$ = 0,1,2,3) to a superspace  with coordinates $Z^M(\sigma) = (x^m(\sigma), \theta^{\beta I}(\sigma))$, where $I=1,2$ denotes the two components of the doublet of 16 component Majorana-Weyl spinors $\theta^I$ in the IIB theory, see  \cite{Bergshoeff:1996tu} for details. Instead of the action \rf{bos} where $\theta^{\beta I}(\sigma)=0$ the complete classical D3 brane action depends on bosonic  and on fermionic  fields on the brane,  $x^m(\sigma)$ and $\theta^{\beta I}(\sigma)$, and on a worldvolume vector field $A_\mu $. It is given by the DBI and WZ expressions in the background superspace
\be
S^q=- T_3 \int d^4 \sigma\, \sqrt {-g_{\mu\nu}+{\cal F}_{\mu\nu}}     + q \,T_3 \, \int C e^{{\cal F}}\,,
\ee
where the pull-back to the worldvolume of the metric is
\be
g_{\mu\nu}\Big (x^m(\sigma), \theta^{\beta I} (\sigma)\Big )= E^a_\mu(x, \theta)  E^b_\nu (x, \theta)\,   \eta_{ab} \, ,
\ee
with $\eta_{ab}$ being the 10d flat Minkowski metric and $E^a_\mu(x, \theta) = \partial_\mu Z^M E_M{}^a(x,\theta)$. The 2-form field strength ${\cal F}= dA- B_2$  has a part $dA$, which is the field strength of the vector field $A_\mu$ on the brane, and $B_2$ is the pull-back to the worldvolume of a 2-form potential $B_2$ in superspace.

The formal sum of the RR-forms $C$ in the WZ term $\int C e^{{\cal F}}$ is integrated over the worldvolume of the brane, which picks out the $4$-form part. This formal sum of RR forms is a form in superspace,
\be
C= \sum_{r \text{ even}} C_r= \sum_{r \text{ even}} {1\over r!} dZ^{M_1} ...dZ^{M_r} C_{M_1...M_r} (x, \theta)\,.
\ee
The corresponding sum of RR field strengths is given by
\be \label{RRcurvatures}
F = dC- H\wedge C\,,
\ee
with $H=dB_2$. The DBI and the WZ terms each separately preserve the unbroken supersymmetry of the underlying superspace with $(x, \theta)$ coordinates.
The local fermionic $\kappa$-symmetry of the total action requires that the variation of the DBI action is canceled by the variation of the WZ action. The $\kappa$-symmetry transformations are defined as follows, with $\delta E^A\equiv \delta Z^M E_M^A$: the part of the supervielbein with a bosonic tangent space index $E^a(x, \theta) = d\sigma^\mu  \partial_\mu Z^M E_M{}^a = d Z^M E_M{}^a (Z)$, does not transform, but the fermionic one does
\be
\delta_{\kappa} E^a=0\,.
\label{kappaEa}
\ee
The fermionic component of the tangent space supervielbein  $ E^{\alpha I}= d\sigma^\mu  \partial_\mu Z^M E_M{}^{\alpha I}$ transforms as follows
\be
\delta_{\kappa} E^{\alpha I}= (\bar \kappa (1+\Gamma))^{\alpha I} \, , \qquad \Gamma^2=1\, , \qquad {\rm tr}\,  \Gamma=0\,,
\label{kappa}
\ee
where $\Gamma$ is a matrix in spinor space  that has an expression in terms of  the super embedding coordinates and the supervielbein, which we refrain from giving here. The transformation of the vector field $A_\mu $ is determined from (\ref{kappaEa}) and (\ref{kappa}) and the components
\begin{equation}
  \delta _\kappa A_\mu =E_\mu {}^A\delta _\kappa E^B\, B_{AB}\,.
 \label{delkappaAmu}
\end{equation}
The variation of the DBI and WZ term is
\be
\delta_\kappa  S_{DBI}= - \int d^4 \sigma {\cal L}_{DBI} \,    \delta_\kappa  E \, N \, , \qquad \delta_\kappa  S_{WZ}= - \int d^4 \sigma {\cal L}_{DBI} \,    \delta_\kappa  E \, P\,,
\ee
where $N, P$ are some complicated functionals of the superfields of the theory and
where $(1+\Gamma) (N+ P)=0$ so that $(N+ P)= (1-\Gamma) T$ for some $T$ and the full action is invariant
\be
\delta_\kappa ( S_{DBI}+ S_{WZ}) =- \int d^4 \sigma {\cal L}_{DBI}   \,  \bar \kappa (1+\Gamma) (1-\Gamma) \, T=0\,.
\ee

\subsection{\texorpdfstring{$\overline {\rm \bf D3}$}{Bar D3} and D3 in a flat superspace background, with orientifolding}

Here we give a short review of \cite{Kallosh:2014wsa}.
In this case, the dependence of all superfields above on  $\theta^I(\sigma)$ is relatively simple \footnote{The index $I=1,2$ on the spinor doublet is contracted with Pauli matrices $(\sigma_n)^I{}_{J}$ or the identity matrix $\unity^I{}_{J}$. To avoid cluttering we often suppress the identity matrix and the indices  $I,J$, etc.. As usual, we also often suppress the spinorial indices like $\alpha, \beta$.}, namely for $a=0,1,2,3$\be
E^a( \theta)= \delta ^a_m dx^m + \bar \theta \Gamma^a d\theta\,, \qquad E^{\alpha I} = \delta^\alpha _\beta d\theta^{\beta I} \,,\qquad B_2= C=0\,.
\ee
In \cite{Kallosh:2014wsa} we have imposed the following orientifold projection condition on the fermions living on the $\overline {\rm D3}$ brane
\be
(\unity - \rmi \sigma_2 \Gamma^{0123})\theta=0\qquad \Leftrightarrow \qquad \theta^2  =  {\Gamma}_{0123} \theta^1  \,,
\label{Orientifold}
\ee
following the analogous $\kappa$-symmetry gauge-fixing condition for the D3 brane in \cite{Bergshoeff:2005yp}. After this orientifold truncation the $\kappa$-symmetry disappears (see \cite{Bergshoeff:1999bx}). The above condition was supplemented by the requirement of a consistent supersymmetric truncation/orientifold condition for the bosons
\be\label{eq:truncation}
{\cal F}_{\mu\nu}=0\,,  \qquad     \Pi_\mu^{\tilde a} = \partial_\mu\phi^{\tilde a}-\bar\theta \Gamma^{\tilde a} \partial_\mu \theta =0\, , \quad \text{with} \quad \tilde a=4,5,\ldots,9\,,
\ee
which corresponds to placing the 3-brane on top of an O3 orientifold plane, which removes all bosonic worldvolume degrees of freedom.
The 1-forms $E^{a}(\theta)=\delta ^a_m dx^{m}+ \bar \theta^1 \Gamma^{a} d\theta^1 + \bar \theta^2 \Gamma^{a} d\theta^2$ are under these restrictions equal to
\be\label{eq:Es}
E^{a}( \lambda)= \delta^a_m dx^{m}+ \bar \lambda \Gamma^{a} d\lambda\, , \qquad \lambda = \sqrt 2 \, \theta^1 = -\sqrt{2} \Gamma_{0123} \theta^2\,.
\ee
Here $\lambda $ is a 16-component 10d Majorana-Weyl spinor.
With all constraints  taken into account  the DBI action of the $\overline {\rm D3}$ takes the form
\be
\label{actionDBI-Restr}
S_{\rm DBI}^{\overline {\rm  D3}}|_{\theta^2-\Gamma_{0123} \theta^1={\cal F}_{\mu\nu}=\Pi_\mu^{\tilde a}=0} =  - T_3\,
\int \rmd^{4} \sigma\, \sqrt{- \det g_{\mu\nu} } = - T_3\, \int d^4\sigma \det E\,.
\end{equation}
The fact that the DBI action reduces in this limit to the VA action has been known for a long time \cite{Kallosh:1997aw} and recently confirmed in \cite{Bergshoeff:2013pia}. The WZ term of the $\overline {\rm  D3}$ brane was shown in \cite{Kallosh:2014wsa} under the conditions \rf{Orientifold}, \rf{eq:truncation} to be equal to
\be
\label{actionWZ-Restr}
S_{\rm WZ}^{\overline {\rm  D3}}|_{\theta^2-\Gamma_{0123} \theta^1={\cal F}_{\mu\nu}=\Pi_\mu^{\tilde a}=0}=  - T_3\, \int d^4\sigma\det E\,,
\end{equation}
and the total action is
\be
\label{actionTotal-Restr}
S_{\rm DBI+WZ}^{\overline {\rm  D3}}|_{\theta^2-\Gamma_{0123} \theta^1={\cal F}_{\mu\nu}=\Pi_\mu^{\tilde a}=0}=  - 2 T_3\, \int d^4\sigma\det E\,,
\end{equation}
based on the brane actions in \cite{Aganagic:1996nn}.
When the fermions $\lambda$ are absent we find that this expression for the $\overline {\rm  D3}$ brane, apart from the warping, is the same as in \cite{Kachru:2003aw}, as  derived in detail in \cite{Kachru:2003sx} and presented above in \rf{CancDoubl}. In our flat superspace background $\lambda$ is a 16-component spinor and the resulting 4d action has an $\mathcal{N}=4$ Volkov-Akulov spontaneously broken supersymmetry.

For the D3 brane in a flat superspace background, with orientifolding \rf{Orientifold}, \rf{eq:truncation}
we find a cancelation between the DBI and WZ term so that the action vanishes
\be
\label{actionTotal-RestrD3}
S_{\rm DBI+WZ}^{ {\rm  D3}}|_{\theta^2-\Gamma_{0123} \theta^1={\cal F}_{\mu\nu}=\Pi_\mu^{\tilde a}=0}=  -  T_3\, \int d^4\sigma\det E \, +  \, T_3\, \int d^4\sigma\det E=0\,.
\end{equation}
This is consistent with the fact that for a D3 brane sitting on top of an O3 plane, all worldvolume fields are projected out, while for a $\overline {\rm  D3}$ brane the scalars are projected out and the 16 component worldvolume fermion remains (see for example \cite{Sugimoto:1999tx,Uranga:1999ib}).

This sixteen component spinor $\lambda(\sigma)$ in equation \eqref{actionTotal-Restr} (cf. equation \eqref{eq:Es}) may be decomposed into four 4d spinors $\lambda^0(\sigma)$, $\lambda^i(\sigma)$, $i=1,2,3$ that transform as the ${\bf 1}$ and {\bf 3} under the $SU(3) \subset SU(4) = SO(6)$ holonomy group of the transverse space (similarly to the vector field $A_\mu$ which is a singlet and the complex scalars $\varphi^i(\sigma)= \frac{1}{\sqrt{2}}\left(\phi^{i+3} + \rmi \phi^{i+6}\right)$ that transform as a triplet). We have argued in \cite{Kallosh:2014wsa} that for the preservation of only the $\mathcal{N}=1$ non-linearly realized supersymmetry one has to truncate the spinor triplet $\lambda^i$, so that the remaining fermion on the brane is the $SU(3)$ singlet
\be
S_{\rm DBI+WZ}^{\overline {\rm  D3}}|_{\theta^2-\Gamma_{0123} \theta^1={\cal F}_{\mu\nu}=\Pi_
\mu^{\tilde a}=\lambda^i=0}=  - 2 T_3\, \int d^4\sigma\det E(\lambda^0)\,, \qquad E^{a}( \lambda^0)= \delta^a_m dx^{m}+ \bar \lambda^0 \gamma^{a} d\lambda^0\,.
\end{equation}
There was the expectation that when we will be able to study the $\overline {\rm D3}$ brane in a GKP  background, that preserves $\mathcal{N}=1$ supersymmetry, then this truncation might be realized in a more clear way. This is indeed the case as we explain in the next section.

\section{\texorpdfstring{$\overline {\rm \bf D3}$}{Bar D3} and D3 in a curved superspace background with unbroken \texorpdfstring{$\mathcal{N}=1$}{N=1} supersymmetry}
\label{ss:bD3D3superspace}

The curved space background is introduced using superfields for all bosonic curved background expressions. For example, for the supervielbein:
\be
E^a( x, \theta)=  e ^a_m(x, \theta)  dx^m +e^a_{\beta I} (x, \theta ) d\theta^{\beta I}\,,
\ee
where $e ^a_m(x, \theta) $ and $e^a_{\beta I} (x, \theta )$ are expanded in powers of $\theta$. The same applies to the form fields $B_2(x, \theta), C_{2}(x, \theta)$ etc. See for example the detailed expressions in type IIB theory required for the quadratic in fermions action for the D3 and $\overline {\rm D3}$ brane in
\cite{Grana, Grana:2003ek,Sandip} in a gauge where $\theta^1= a \Theta$ and $\theta^2= b\Theta$ with $a^2+b^2=1$. In the gauge \rf{Orientifold} $\theta^1$ and $\theta^2$ are not proportional, and our gauge therefore does not belong to that class of gauges.
However, we will use below the construction of the classical (not gauge-fixed) action quadratic in $\theta$ for the $\overline {\rm  D3}$ and D3 brane, based on the superspace of $d=11$ supergravity for the M2 brane background and a T-duality as in \cite{Marolf:2003ye,Toine, Bergshoeff:2005yp}.

Before we will look at the important details involving the properties of the curved background with fluxes, we would like to make the observation that the generalization of the $\overline {\rm  D3}$ brane action \rf{actionTotal-Restr} and of the D3 brane action \rf{actionTotal-RestrD3} to the curved background with unbroken $\mathcal{N}=1$ supersymmetry requires the computation of the action where the supervielbein is defined in a curved superspace:
\be
S_{\rm DBI+WZ}^{\overline {\rm  D3}}\Rightarrow   - 2 T_3\, \int d^4\sigma\det E \,, \qquad E^{a}= e^a_m( x, \theta) dx^{m}+ e^a_{\beta I} ( x, \theta)d\theta^{\beta I}\,,
\end{equation}
\be
S_{\rm DBI+WZ}^{\rm  D3}\Rightarrow   -  T_3\, \int d^4\sigma\det E \, +  T_3\, \int d^4\sigma\det E=0\,.
\end{equation}
This seems to be a straightforward generalization of the flat background results discussed in the previous section and is indeed what we find from an explicit calculation to quadratic order in the fermions.

As mentioned above, in the classical action $\theta$ is a doublet of $d=10$ Majorana-Weyl spinors. Upon gauge-fixing the $\kappa$-symmetry, the remaining spinor is one $d=10$ Majorana-Weyl spinor. For a transverse space with $SU(3)$ holonomy, we can split this 16-component spinor into four 4d spinors that transform as a singlet and triplet under $SU(3)$. We would like to investigate the properties of these fields living on a spacetime filling $\overline {\rm  D3}$ brane probing a supersymmetric GKP background with ISD fluxes. For this purpose we look more carefully at the quadratic in $\theta$ approximation of the brane actions.

We would like to keep the following issue in mind: in the flat background we have found it useful in \cite{Kallosh:2014wsa} to impose the orientifold condition on spinors \rf{Orientifold} and bosons \rf{eq:truncation}, to obtain a consistent supersymmetric truncation of the D3 and $\overline {\rm  D3}$ brane actions. Such a truncation corresponds to a D3 or $\overline {\rm  D3}$ brane on top of an O3 plane.
In the presence of the curved background and fluxes preserving $\mathcal{N}=1$ supersymmetry, we have to revisit this issue. ISD flux carries D3 brane charge. If we have a compact $CY_3$ manifold, then we cancel this charge using O3 planes. We can add a $\overline {\rm  D3}$ brane on top of any of the O3 planes and the projection condition will be identical to the flat space case considered in \cite{Kallosh:2014wsa}: Only the 16 component spinor survives. We will study the corresponding fermionic action in the next two subsections.

\subsection{The D3 brane and \texorpdfstring{$\overline {\rm \bf D3}$}{bar D3} brane actions quadratic in fermions}\label{sec:quadratic}

The classical D3 brane action quadratic in $\theta$ in string frame is given by \cite{Marolf:2003ye, Toine} and \cite{Bergshoeff:2005yp}
\be
{\cal L}_f^{\rm D3}= {1\over 2} e^{-\phi} T_3 \sqrt {-\det g} \, \bar \theta \, (1-\Gamma_{\rm D3}) \, [ \Gamma^\mu \delta \psi_\mu -\delta \lambda] \theta\,,\qquad
\Gamma_{\rm D3}= \rmi \sigma_2  \frac{1}{\sqrt{-g}}  {\Gamma}_{0123}\,,\qquad \Gamma^m =  e^m_a \tilde{\Gamma}^a\,,
\ee
where $\Gamma_{\rm D3}$ is the $\kappa$-symmetry operator in \rf{kappa},  $\delta \bar \theta = \bar \kappa (1-\Gamma_{\rm D3})$, the $\Gamma$'s are the 10-dimensional gamma-matrices pulled back to the brane  and the $\tilde{\Gamma}$ denote the flat space gamma-matrices.
$\delta \psi_\alpha$ and $\delta \lambda$ are expressions for the local supersymmetry variation of the gravitino and dilatino in IIB supergravity and here we follow \cite{Bergshoeff:2005yp}. The dependence on ${\cal F}$ is omitted here.

The classical action of the $\overline {\rm  D3}$ brane is
\be\label{eq:D3baraction}
{\cal L}_f^{\overline {\rm D3}}= {1\over 2} e^{-\phi} T_3 \sqrt {-\det g} \, \bar \theta \, (1-\Gamma_{\overline{\rm D3}}) \, [ \Gamma^\mu \delta \psi_\mu -\delta \lambda] \theta\,,\qquad
\Gamma_{\overline{\rm D3}}= - \Gamma_{\rm D3}\,.
\ee
In detail we have
\be
{\cal L}_f^{\overline {\rm D3}} = \frac12 e^{-\phi} T_3 \sqrt{- \det g}\, \bar{\theta} \, (1+\Gamma_{\rm D3})\, [\Gamma^\mu \nabla_\mu +U_1 +  U_3 + U_5 ] \theta\,,
\ee
with
\bea\label{eq:Us}
U_1 &=& -\frac12 \Gamma^m \partial_m \phi + \frac14 e^\phi F_\mu \Gamma^\mu (\rmi \sigma_2)\,,\cr
U_3 &=& \frac18 (H_{\mu n p} \sigma_3 + e^{\phi} F_{\mu n p} \sigma_1) \Gamma^{\mu n p} - \frac{1}{24} (H_{mnp} \sigma_3 + e^\phi F_{mnp}\sigma_1)\Gamma^{mnp}\,,\cr
U_5 &=&\frac{1}{8 \cdot 4!} e^\phi F_{\mu n p q r} \Gamma^{\mu n p q r} (\rmi \sigma_2)\,.
\eea
Here $\mu=0,1,2,3$ stands for the world-volume directions of the brane, $m,n,p$ denote all ten spacetime directions and $F_\mu$ is one of the RR curvatures defined in eq.~(\ref{RRcurvatures}).

We are now ready to present one of the main observations of this paper, namely that  {\it the fermionic action at the level quadratic in $\theta$ shows a cancelation for the D3 brane and a `doubling' for the $\overline {\rm D3}$ brane, independent of the presence or absence of the curved background and fluxes}. Namely, once we impose the orientifold condition on spinors, the D3 brane action vanishes and the $\overline {\rm  D3}$ brane action `doubles'
\be \label{eq:gauge}
\bar \theta \, (1-\Gamma_{{\rm D3}})=0\, \qquad \Rightarrow \qquad {\cal L}_f^{\rm D3} =0\, ,\qquad {\cal L}_f^{\overline {\rm D3}}= e^{-\phi} T_3 \sqrt {-\det g} \, \bar \theta  \, [ \Gamma^\mu \delta \psi_\mu - \delta \lambda] \theta\,,
\ee
as we have seen in the flat background in \cite{Kallosh:2014wsa} with ${\cal L}_f^{\rm D3} =0$ and ${\cal L}_f^{\overline {\rm D3}}= T_3 \, \bar \theta  {\slashed{\partial  }}  \theta$. In the presence of the curved background and fluxes or in their absence, the kinetic term for the fermions vanishes for the D3 brane. This means that all bosonic fields on the brane have to vanish for consistency of the supersymmetric action with ${\cal L}_f^{\rm D3} =0$. This is the condition \rf{eq:truncation} which we have imposed in the case of the flat background and which corresponds to the O3 plane projection conditions for the worldvolume fields, that remove for a D3 brane all bosons and fermions.

In case of the curved background we would like to study the action of the ${\overline {\rm  D3}}$ brane in the GKP  $\mathcal{N}=1$ supersymmetry preserving background. Such an action was studied in \cite{McGuirk:2012sb} in the gauge $\theta^2=0$.  It was found there how the masses of worldvolume scalars and fermions depend on the background. We will find that our ${\overline {\rm  D3}}$ brane with the orientifolding condition (\ref{Orientifold}) imposed on fermions has the same mass terms for the fermions since this answer is independent of the particular gauge fixing.

\subsection{The \texorpdfstring{$\overline {\rm \bf D3}$}{bar D3} brane action in a GKP background}

We are interested in a spacetime filling $\overline {\rm D3}$ brane in a GKP type background \cite{Giddings:2001yu}. We restrict ourselves to the particular truncation of the fermions
\be\label{eq:gauge2}
\bar \theta \, (1-\Gamma_{{\rm D3}})=0 \qquad  \Leftrightarrow \qquad \theta^2  =  \tilde {\Gamma}_{0123} \theta^1  \,,
\ee
that is consistent with an O3 orientifold projection. If we place a single D3 brane on top of an O3 plane then all worldvolume degrees of freedom are projected out, which is consistent with the vanishing of the action that we found in \eqref{eq:gauge}. For an $\overline {\rm  D3}$ on top of an O3 plane, the vector and scalars are truncated and we are left with the fermionic degrees of freedom that are contained in the 10d Majorana-Weyl spinor $\theta^1$. In order to evaluate the $\overline {\rm  D3}$ action, we recall that for any 10d Majorana-Weyl spinor
\be\label{eq:contraction}
\bar{\theta}^1 \Gamma^{N_1 N_2 \ldots N_n} \theta^1 =0\,, \quad  \text{for} \quad n \in \{0,1,2,4,5,6,8,9,10\}\,.
\ee
This fact leads to a substantial simplification of the 3-brane action in a GKP background \cite{Giddings:2001yu}. The axio-dilaton $\tau = C_0 + \rmi e^{-\phi}$ in such a background can vary holomorphically along the internal dimensions transverse to 7-brane sources. However, the corresponding dilaton term in $U_1$ in \eqref{eq:Us} does not contribute to the action due to \eqref{eq:contraction}. The second term in $U_1$ in \eqref{eq:Us} does not contribute to the action either since $F_\mu=\partial_\mu C_0 = 0$ for $\mu=0,1,2,3$ in the GKP background. The non-vanishing 5-form flux has either zero or four legs\footnote{With four legs along the brane the term is proportional to $\bar \theta^1 \Gamma^{0123}\Gamma^r\theta^2$, which, using \eqref{eq:gauge2}, is proportional to
$\bar \theta^1\Gamma^r\theta^1=0$.} along the $\overline {\rm  D3}$ brane. Therefore the $U_5$ contribution as given in \eqref{eq:Us} vanishes as well when contracted with the spinor. Lastly, the 3-form fluxes have no legs along the four non-compact spacetime directions, i.e. along the $\overline {\rm  D3}$ brane: $H_{mnp}$ and $F_{mnp}$ are only non-zero when $m,n,p \in \{4,5,6,7,8,9\}$. Thus the action in string frame in a general GKP background and in the gauge \eqref{eq:gauge2} takes the simple form
\be\label{eq:D3action1}
{\cal L}_f^{\overline {\rm D3}} = T_3 e^{4 A_0} \, \bar{\theta}^1 [2 e^{-\phi} \Gamma^\mu \nabla_\mu - \frac{1}{12} (e^{-\phi} H_{mnp}  + F_{mnp} \tilde{\Gamma}_{0123}) \Gamma^{mnp}] \theta^1\,,
\ee
where $A_0$ denotes the warp factor evaluated at the position of the $\overline{\rm D3}$ brane, see (\ref{metricGKP}-\ref{pot2}).

We now use the fact that our 10d Majorana-Weyl spinor $\theta^1$ satisfies $\theta^1 = \tilde{\Gamma}^{0123456789} \theta^1$ which implies $\tilde{\Gamma}_{0123} \theta^1 = \tilde{\Gamma}_{456789} \theta^1$. Using this, we can rewrite the last term in \eqref{eq:D3action1} as follows
\begin{eqnarray}
F_{mnp} \tilde{\Gamma}_{0123} \Gamma^{mnp}\theta^1&=& F_{mnp}\,e^m_ae^n_be^p_c\tilde \Gamma^{abc}\tilde{\Gamma}_{456789}\theta^1= \frac{1}{3!}F_{mnp}\,e^m_ae^n_be^p_c\tilde \Gamma_{def}\varepsilon^{defabc}\theta^1\nonumber\\
&=&\frac{1}{3!\sqrt{g}}F_{mnp}\varepsilon^{qrsmnp}\Gamma_{qrs}\theta^1= (*_6 F_{(3)})^{mnp} \Gamma_{mnp}\theta^1\,,\nonumber\\
 (*_6 F_{(3)})^{mnp} & \equiv  &\frac{1}{3!\sqrt{g}}\varepsilon^{mnpqrs}F_{qrs}\,,
 \label{Hodgecalculation}
\end{eqnarray}
where $*_6$ denotes the internal six dimensional Hodge dual. We can thus write the action as
\be\label{eq:antiD3action}
{\cal L}_f^{\overline {\rm D3}} = T_3 e^{4 A_0} \, \bar{\theta}^1 [2 e^{-\phi} \Gamma^\mu \nabla_\mu - \frac{1}{12} (e^{-\phi} H_{mnp} + (*_6 F_{(3)})_{mnp}) \Gamma^{mnp}] \theta^1\,.
\ee
We now express the action in terms of the more familiar complexified 3-form flux $G_{(3)} = F_{(3)} - \tau H_{(3)}$. The $G_{(3)}$ flux can be uniquely decomposed into an imaginary self-dual (ISD) and imaginary anti-self-dual (IASD) part
\be\label{eq:ISD}
G_{(3)} = G_{(3)}^{\text{ ISD}} + G_{(3)}^{\text{IASD}}\,, \qquad G_{(3)}^{\text{ ISD}} = \frac12 (G_{(3)}-\rmi*_6 G_{(3)})\,, \qquad G_{(3)}^{\text{IASD}} = \frac12 (G_{(3)} + \rmi*_6 G_{(3)})\,.
\ee
Assuming that the pull-back of $C_0$ to the brane vanishes we find the simple relation
\be
e^{-\phi} H_{(3)} = \frac{\rmi}{2} \lp G_{(3)} -\bar{G}_{(3)} \rp\,, \qquad F_{(3)} = \frac12 \lp G_{(3)} +\bar{G}_{(3)} \rp\,,
\ee
where the complex conjugates are denoted as
\begin{equation}
  \bar{G}_{(3)} = F_{(3)} + \rmi e^{-\phi} H_{(3)}\,,\qquad \bar{G}_{(3)}^{\text{ISD}}= \frac12\left(\bar{G}_{(3)}+\rmi*_6 \bar{G}_{(3)}\right)\,.
 \label{barGISD}
\end{equation}
Plugging this into the $\overline {\rm  D3}$ action \eqref{eq:antiD3action} and using \eqref{eq:ISD} we find
\be\label{eq:antiD3action2}
{\cal L}_f^{\overline {\rm D3}} = T_3 e^{4 A_0} \, \bar{\theta}^1 \ls 2 e^{-\phi} \Gamma^\mu \nabla_\mu -\frac{\rmi}{12}\lp G^{{\rm ISD}}_{mnp} - \bar{G}^{{\rm ISD}}_{mnp} \rp \Gamma^{mnp}\rs \theta^1\,.
\ee
Note that, as expected, the ${\overline {\rm D3}}$ brane couples only to the ISD part of the $G_{(3)}$ flux and the IASD part drops out of the action. The analogue result was obtained for a D3 brane in ISD flux in eq. (36) of \cite{Grana}, see also \cite{McGuirk:2012sb}. In particular, up to an overall factor, we find that the first line in equation (36) in \cite{Grana} is for $a=1,b=0$, i.e. in the $\theta_2=0$ gauge, the same as $\eqref{eq:antiD3action2}$ with IASD flux instead of the ISD flux. This is expected since the particular gauge fixing should not change the action, and the D3 brane and ${\overline {\rm D3}}$ brane actions are related by a sign flip of the RR fields, which is equivalent to the sign flip $\Gamma_{\overline {\rm D3}} = -\Gamma_{{\rm D3}}$ (cf. \eqref{eq:D3baraction}), and which maps $G_{(3)}^{\text{ISD}}$ to $-G_{(3)}^{\text{IASD}}$ flux and vice versa.

The 16-component spinor $\theta^1$ can be decomposed into four 4d Dirac spinors $\lambda^0$, $\lambda^i$ with $i=1,2,3$. $\lambda^0$ is a singlet under the $SU(3)$ holonomy group of the internal Calabi-Yau manifold while the $\lambda^i$ transform as a triplet. We use $\pm$ subscripts to denote 4d Weyl spinors that satisfy $\lambda_\pm = \frac12(1 \pm \rmi   {\tilde{\Gamma}}_{0123}) \lambda$. These notations are explained in detail in Appendix \ref{app:d10d4}.
Following section 2 and 6 of \cite{Grana} (see also our Appendix \ref{app:example}, where we work out an explicit example), we can write the $\overline {\rm D3}$ brane action in terms of the corresponding 4d Weyl spinors:
\begin{eqnarray}
{\cal L}_f^{\overline {\rm D3}} &=&2 T_3 e^{4 A_0-\phi} \, \left[\bar \lambda_-^\zm\gamma^\mu\nabla _\mu\lambda_+^\zp+
  \bar \lambda_-^{\jm}\gamma^\mu\nabla _\mu\lambda_+^{\ip}\delta_{\ip\jm}\right. \label{eq:antiD3action4d}\\
 &&\left. \phantom{2 T_3 }+\ft12m_0 \bar \lambda_+^\zp\lambda_+^\zp +\ft12\overline{m}_0 \bar \lambda_-^{\zm}\lambda_-^{\zm}
  +m_i\bar \lambda_+^{\zp}\lambda_+^{\ip}+\overline{m}_{\ib}\bar\lambda_-^{\zm} \lambda_-^{\im}+\ft12 m_{ij}\bar \lambda_+^{\ip}\lambda_+^{\jp}+\ft12\overline{m}_{\ib\jb}\bar \lambda_-^{\im}\lambda_-^{\jm}\right]\,,
\nonumber
\end{eqnarray}
where
\begin{eqnarray}
 m_0 & = &\frac{\sqrt{2}}{12}\rmi e^\phi \bar \Omega^{uvw}\bar G^{{\rm ISD}}_{ u v  w}\,,\qquad \qquad\qquad \qquad  \qquad \qquad \qquad  \mbox{from }(0,3)\mbox{ flux,}\label{eq:msing}\\
 m_i & = & -\frac{\sqrt{2}}{4}e^\phi e_i^u\,\bar G^{{\rm ISD}}_{uv\bar w} J^{v\bar w}\,,\qquad \qquad \qquad \qquad \mbox{from non-primitive }(1,2)\mbox{ flux,}\label{eq:mmix}\\
 m_{ij}&= &\frac{\sqrt{2}}{8}\rmi e^\phi\left(e_i^{w}e_j^t+e_j^{w}e_i^t\right)\Omega_{uvw}g^{u\bar u}g^{v\bar v}\bar G^{{\rm ISD}}_{t \bar u\bar v}\,,\quad\mbox{ from primitive }(2,1)\mbox{ flux.}
 \label{eq:mtrip}
\end{eqnarray}
The $\overline{m}$'s are the complex conjugate of the $m$'s and the \K\ form $J$ and holomorphic 3-form $\Omega$ arise when going to curved indices (see appendix \ref{app:d10d4} for details). The masses of the fermions on an ${\overline {\rm D3}}$ brane in an ISD background were studied in \cite{McGuirk:2012sb} in the fermionic $\kappa$-symmetry gauge  $\theta^2= 0$,  in the Einstein frame. Our results for the masses are consistent with theirs, as shown in eqs.  (3.99a)-(3.99c) in \cite{McGuirk:2012sb}, as expected from the gauge-independence of physical observables.

We are interested in the case when the ISD flux preserves $\mathcal{N} =1$ supersymmetry so that supersymmetry is only broken spontaneously by the ${\overline {\rm D3}}$ brane. Using the notation $(p,q)$ to denote $p$ holomorphic and $q$ anti-holomorphic indices, see (\ref{ISDcomponents}), this means that the ISD flux has only non-zero  $(2,1)$ components $G_{uv\bar w}\neq 0$ and is primitive, i.e. the contraction with the \K\, form vanishes $G_{uv\bar w}J^{v\bar w}=\rmi{G_{uv}}^v=0$, \cite{Grana:2001xn}. In particular, we are interested in the case in which the $(0,3)$ part $G_{\bar u \bar v \bar w}=0$ is absent.

Due to the properties of the ISD fluxes with vanishing $(0,3)$ part, $G_{\bar u\bar v \bar w}= 0$, the corresponding mass of the singlet $\lambda^0$ in (\ref{eq:msing}) vanishes. Due to the primitivity of the $(2,1)$ component, ${G_{uv}}^v=0$, and the absence of $(1,2)$ flux, the mixing terms between the singlet and the triplet vanish as well in \rf{eq:mmix}. The fermion triplet has a mass that is set by the ISD fluxes as shown in \eqref{eq:mtrip}:
\begin{equation}
  m_0=m_i=0 \, \qquad m_{ij}\neq 0\,.
 \label{massesSUSY}
\end{equation}
The triplet can therefore be integrated out and does not appear in the low energy effective action. All of this is self-consistent: the ${\overline {\rm D3}}$ brane breaks  $\mathcal{N}=1$ supersymmetry spontaneously. The singlet fermion $\lambda^0$ is the Goldstino and remains massless  and does not mix with the triplet $\lambda^i$.
Note, that one can modify the background by turning on (0,3) ISD flux.  In this case the background breaks all supersymmetries and the singlet $\lambda^0$ gets a mass (cf. \eqref{eq:msing}). This is  the situation where the (0,3) ISD flux in the bulk generates the superpotential, $W= \int G_3 \wedge \Omega$,  and the supersymmetry of the effective supergravity is broken spontaneously.

If we would not have placed the ${\overline {\rm D3}}$ brane on top of an orientifold plane, then additional bosonic degrees of freedom would have been present on the brane. The three complex scalars that control the position of the ${\overline {\rm D3}}$ brane in the internal CY$_3$ manifold get a mass in the GKP background. In particular, this mass of the complex scalars $\varphi^i$ living on an $\overline {\rm  D3}$ brane in a supersymmetric ISD background was computed in \cite{McGuirk:2012sb} and was found to be proportional to
\be
m_{i\jb} \propto \partial _i \partial _{\jb} e^{A(z, \bar z)}\big|_{z=z_0}\,.
\ee
It is non-vanishing and not related to the mass of the fermion triplet which is determined by the $(2,1)$ component of the ISD flux $G_{(3)}$. Again this is consistent with our finding that supersymmetry is spontaneously broken. At low energies the scalars and the fermion triplet can therefore be integrated out, leaving the massless goldstino $\lambda^0$.

\section{Discussion}
\label{ss:discussion}

In this paper we studied the action of a single $\overline {\rm  D3}$ brane in a GKP background \cite{Giddings:2001yu}, preserving $\mathcal{N} =1$ supersymmetry. We are   using the orientifolding condition imposed on fermions, $\theta^2= \tilde{\Gamma}_{0123} \theta^1$, with and without truncation from the spectrum the worldvolume scalar fields and the vector. Such a truncation of the bosonic degrees of freedom was suggested by the fact that the D3 brane under the same orientifold condition on the fermions does not have a kinetic term for the fermions, hence all bosons have to be cut off, both on the D3 brane as well as on the $\overline {\rm  D3}$ brane. This truncation corresponds, in string theory terms, to placing the 3-brane on top of an O3 plane. The 10d worldvolume spinor $\theta^2= \tilde{\Gamma}_{0123} \theta^1$ can be decomposed into four 4d Dirac spinors that transform as a singlet and triplet under the $SU(3)$ holonomy group of the internal Calabi-Yau manifold. We have found in the truncated case, that the spinor triplet on the $\overline {\rm D3}$ brane is generically massive in a GKP background due to the ISD flux. The additional $SU(3)$ singlet does not mix with the triplet and remains massless, iff the GKP background preserves $\mathcal{N}=1$ supersymmetry. The fact that the background fluxes give a mass to the fermion triplet is consistent with our orientifolding  condition, which even in the flat background \cite{Kallosh:2014wsa} was truncating the triplet of complex scalars from the spectrum. Thus in the presence of the background we have a fully consistent argument that the $\overline {\rm  D3}$ brane breaks supersymmetry spontaneously and the remaining fermion singlet is a goldstino. It is therefore convincing that the  effective action on the $\overline {\rm  D3}$  brane is the VA action for this singlet fermion. This then establishes the existence of a 4d nilpotent chiral superfield describing the low energy effective action of a $\overline {\rm D3}$ brane in the $\mathcal{N}=1$ supersymmetry preserving GKP background.

If we do not truncate the worldvolume bosons on the $\overline {\rm  D3}$ brane one finds that the mass of the scalars on the $\overline {\rm  D3}$ brane, according to \cite{McGuirk:2012sb}, is also non-vanishing in the GKP background, and it is different from the masses for the fermion triplet. In this case we might say that for the $\overline {\rm  D3}$ brane the triplet of massive $\mathcal{N}=1$ multiplets is either truncated, or might be represented in the effective theory as a multiplet with spontaneous supersymmetry breaking via a superpotential.

We can relate the $\overline {\rm  D3}$ brane action in the GKP background, preserving $\mathcal{N} =1$ supersymmetry to $\mathcal{N}=1$ supergravity with de Sitter vacua and spontaneously broken $\mathcal{N}=1$ supersymmetry of the VA type as follows
\be
W= M^2 S\, , \qquad K=  S\bar S\,  \qquad {\rm at} \qquad S^2=0 \qquad \Rightarrow  \qquad V= M^4\,,
\label{ModelN}
\ee
where $2T_3  e^{4A_0}= M^4$. Here  $S$ is the nilpotent chiral multiplet which provides a manifestly supersymmetric version of the Volkov-Akulov goldstino (see for example \cite{Komargodski:2009rz} in the case of global susy). After computing the scalar potential from the K\"{a}hler and superpotential, we have to set the scalar part of the superfield $S$ to zero. This nilpotent superfield  $S$  originates from the fermion singlet on the $\overline {\rm  D3}$ brane. We may now either argue that a consistent orientifolding involves truncating the three complex scalars and their fermionic partners as well as the vector, or we may say that, in the presence of the GKP background, the scalars and the fermion triplet are heavy and have been integrated out.
Either way, the relation between the fermions living on a $\overline {\rm D3}$ brane and the goldstino multiplet in the effective $\mathcal{N}=1$ supergravity is now clearer than it was before we included the GKP background.

Our supergravity model \rf{ModelN} can be extended to a supergravity version of the KKLT-type model \cite{Kachru:2003aw,Kallosh:2004yh}, proposed in \cite{Ferrara:2014kva}, which in addition to the volume modulus involves the nilpotent superfield $S$. In particular, we imagine that we first stabilize all closed string moduli except the volume modulus in a supersymmetric Minkowksi vacuum using (2,1) ISD flux. Then we add a $\overline {\rm D3}$ brane and two non-perturbative terms involving the volume modulus. Since the $\overline {\rm D3}$ brane and the non-perturbative effects are localized in the internal space, we can take them to be well separated from each other so that it is plausible that the results for the $\overline {\rm D3}$ brane action are unaltered. We then have the following \K \ and superpotential
\be
W= Ae^{-a\rho} -B e^{-b \rho} + M^2 S\, , \qquad K= - 3 \ln(\rho + \overline{\rho})+ S\bar S\,,\qquad {\rm at} \qquad S^2=0 \,.
\label{KKLT}
\ee
When the $\rho$-modulus is stabilized at $\rho=\rho_0$ in an AdS minimum with $D_\rho W=0$, then the potential is
\be
V= \frac{M^4-3|W_0|^2}{(\rho_0+\bar{\rho}_0)^3}\,,   \qquad {\rm at} \qquad S^2=0 \,,
\ee
with $W_0 \equiv W(\rho_0)$. An alternative version would be to take  $K= - 3 \ln(\rho + \overline{\rho}+ S\bar S)$ in which case the uplifting term would be instead $V= \frac{M^4}{(\rho_0+\bar{\rho}_0)^2}$. This would describe the highly warped compactification, as explained around eq. (5.14) in \cite{Kachru:2003sx}.

Thus the result of our studies  of the $\overline {\rm  D3}$ brane in the GKP background is the string theory explanation of the origin of the positive contribution to the energy in de Sitter landscape. In effective $d=4$ supergravity with spontaneously broken $\mathcal{N}=1$ supersymmetry this contribution is given by  ${M^4  \over (\rho_0+\bar{\rho}_0)^n}$ , with $n=3$ (or $n=2$ for highly warped compactification). The parameter $M$ is related to the tension of the $\overline {\rm  D3}$  and the warp factor in the metric. Therefore the uplifting energy can take many different values in the string landscape.

\section*{Acknowledgments}

We are grateful to M. Aganagic, S. Ferrara, M. Gra\~na, S. Kachru, J. Maldacena, L. McAllister, J. Rosseel, J.~Polchinski, T. Van Riet, N. Seiberg, E. Silverstein and S. Trivedi for useful discussions.
The work of KD is supported in part by the Simons Foundation Fellowship and in part by the National Science and Engineering Research Council of Canada.
RK is supported by the SITP and by the NSF Grant PHY-1316699  by the Templeton foundation grant `Quantum Gravity Frontiers'.
AVP is supported in part by the FWO - Vlaanderen, Project No. G.0651.11, and in part by the Interuniversity Attraction Poles Programme initiated by the Belgian Science Policy (P7/37).
This work has been supported in part by COST Action MP1210 `The String Theory Universe'. EB and AVP thank the Department of Physics of Stanford University and the Templeton foundation for the hospitality during a visit in which this work was initiated.
\appendix

\section{Spinor reduction from \texorpdfstring{$d=10$}{d=10} to \texorpdfstring{$d=4$}{d=4}}\label{app:d10d4}

In this appendix we explain how we reduce the expressions of the IIB supergravity to 4 dimensions. We use the gamma matrix and spinor conventions of the book  \cite{Freedman:2012zz}. We use the decomposition of spinor space from 32 components to $4\times 8$ component ones using the $4+6$ gamma matrices\footnote{As in the bulk of the paper we use a tilde to denote the flat space gamma matrices so that we have $\{\Gamma^m,\Gamma^n\} = 2 g^{mn} = e^m_a e^n_b \{\tilde{\Gamma}^a,\tilde{\Gamma}^b\} = 2 e^m_a e^n_b \eta^{ab}$.}
\begin{eqnarray}
  &&\tGamma^a =\tilde{\gamma}^a\otimes \unity _8\,, \qquad a=0,1,2,3\,,\qquad \tGamma^{\tilde{a}} = \gamma_*\otimes \tilde{E}^{\tilde{a}}\,,\qquad \tilde{a}=4,\ldots ,9\,,\nonumber\\
  &&\Gamma_*=-\tGamma_{0\ldots 9}=\gamma_*\otimes E_*\,,\qquad  \gamma_*=\rmi \tilde{\gamma}_{0123}\,,\qquad E_*= \rmi \tilde{E}_{456789}\,.
 \label{Gamma10}
\end{eqnarray}
The $\tilde{E}^{\tilde{a}}$ form a $8\times 8$ Euclidean gamma matrix representation. The charge conjugation matrices satisfy
\begin{equation}
  C_{(10)}=C_{(4)}\otimes C_{(6)} \,,\qquad C_{(10)}^T = -C_{(10)}\,,\qquad C_{(4)}^T = -C_{(4)}\,,\qquad C_{(6)}^T = C_{(6)}\,.
 \label{C1046}
\end{equation}
These have the properties that
\begin{equation}
  (C_{10}\tGamma^a)^T= C_{10}\tGamma^a\,,\qquad (C_{4}\tilde{\gamma}^a)^T= C_{4}\tilde{\gamma}^a\,, \qquad (C_{(6)}\tilde{E}^{\tilde{a}})^T = -C_{(6)}\tilde{E}^{\tilde{a}}\,.
 \label{Cpropgamma}
\end{equation}
Following methods of \cite{Grana}, we introduce the 3 commuting hermitian matrices
\begin{equation}
  S^i = \rmi\tGamma^{(i+3)}\tGamma^{(i+6)}\,,\qquad \mbox{for }i=1,2,3.
 \label{Sidefined}
\end{equation}
These are traceless and square to $\unity $. Hence they have simultaneous eigenvalues $\pm 1$.
They satisfy
\begin{equation}
  S^1S^2S^3=  \rmi \tGamma_{456789}= \unity \otimes E_*\,.
 \label{S123}
\end{equation}

An explicit basis in which the $S^i$ are diagonal is
\begin{eqnarray}
  \tilde{E}_4 & = &  \sigma _1 \otimes \unity \otimes \sigma _3 \,, \nonumber\\
  \tilde{E}_5 & = &  \sigma _3 \otimes \sigma _1 \otimes \unity \,, \nonumber\\
\tilde{E}_6 & = &  \unity \otimes \sigma _3 \otimes \sigma _1 \,, \nonumber\\
\tilde{E}_7 & = &  \sigma _2 \otimes \unity \otimes \sigma _3 \,, \nonumber\\
  \tilde{E}_8 & = &  \sigma _3 \otimes \sigma _2 \otimes \unity \,, \nonumber\\
\tilde{E}_9 & = &  \unity \otimes \sigma _3\otimes  \sigma _2 \,, \nonumber\\
E_* & = & - \sigma _3 \otimes\sigma _3\otimes \sigma_3\,,\nonumber\\
 C_{(6)}&=& \sigma_1\otimes \sigma_1\otimes \sigma_1\,.
\label{ED6}
\end{eqnarray}
In this basis, the three $S^i$ matrices are
\begin{eqnarray}
 S^1 & = & -\,\unity _4\otimes \sigma_3\otimes \unity \otimes \unity \,,  \nonumber\\
 S^2 & = & -\,\unity _4\otimes\unity \otimes \sigma_3\otimes \unity \,,  \nonumber\\
 S^3 & = & -\,\unity _4\otimes\unity \otimes \unity\otimes\sigma_3 \,.
 \label{S123basis}
\end{eqnarray}
We define the complexified matrices ${\tGamma}_{(c)}^i$ and ${\tGamma}_{(c)}^{\ib}$:
\begin{eqnarray}
  &&\tGamma_{(c)}^i = \frac{1}{\sqrt{2}} \left(\tGamma^{i+3} + \rmi \tGamma^{i+6}\right)\,, \qquad \tGamma_{(c)}^{\ib} = \frac{1}{\sqrt{{2}}}\left(\tGamma^{i+3} - \rmi \tGamma^{i+6}\right)\,, \qquad i=1,2,3\,,\nonumber\\
&&\{\tGamma_{(c)}^i,\tGamma_{(c)}^{\jb}\}=2\delta^{i \jb} =2 g^{u \bar v} e_u^i e_{\bar v}^{\jb}\,.
 \label{defGammac}
\end{eqnarray}
In the representation (\ref{ED6}) they are
\begin{eqnarray}
 {\tGamma}_{(c)}^1= \sqrt{2}\gamma_*\otimes \sigma_+\otimes \unity \otimes \sigma _3\,, & \qquad  & {\tGamma}_{(c)}^{\bar{1}}= \sqrt{2}\gamma_*\otimes \sigma_-\otimes \unity \otimes \sigma _3 \,,\nonumber\\
 {\tGamma}_{(c)}^2= \sqrt{2}\gamma_*\otimes \sigma _3 \otimes\sigma_+\otimes \unity\,, & \qquad  & {\tGamma}_{(c)}^{\bar{2}}= \sqrt{2}\gamma_*\otimes \sigma _3\otimes \sigma_-\otimes \unity \,,\nonumber\\
 {\tGamma}_{(c)}^3= \sqrt{2}\gamma_*\otimes \unity \otimes \sigma _3\otimes\sigma_+\,, & \qquad  & {\tGamma}_{(c)}^{\bar{3}}=\sqrt{2}\gamma_*\otimes  \unity \otimes \sigma _3\otimes \sigma_- \,.
 \label{Gammacreps}
\end{eqnarray}
with
\begin{equation}
  \sigma_+=\begin{pmatrix}0&1\cr 0&0\end{pmatrix}\,,\qquad \sigma_-=\begin{pmatrix}0&0\cr 1&0\end{pmatrix}\,.
 \label{sigma+-}
\end{equation}

We now review how one decomposes the 10d Majorana-Weyl spinor $\theta^1= \tGamma^{0123456789} \theta^1= \Gamma_*\theta^1$ into four 4d Weyl spinors $\lambda_\pm^0$ and $\lambda_\pm^i$, with eigenvalues $(\pm 1,\pm 1, \pm 1)$ under the three $S^i$ matrices. We defined in (\ref{eq:Es}):
$\lambda=\sqrt{2}\theta^1$. This we decompose in $\lambda=\lambda_++\lambda_-$ with
\begin{equation}
  \lambda_+= (\gamma_*\otimes \unity )\lambda_+= (\unity \otimes E_*)\lambda_+\,,\qquad \lambda_-=-(\gamma_*\otimes \unity )\lambda_-= -(\unity \otimes E_*)\lambda_-\,.
 \label{lambda+-}
\end{equation}
Denoting the eigenvalues of $S^i$ by $s^i$,  (\ref{S123})  implies that (cf. \cite{Grana})
$\lambda_\pm$ should have $s^1s^2s^3=\pm 1$, i.e. $\lambda_+$ should have three or one positive eigenvalues and  $\lambda_-$ two or zero. Under the $SU(3)$ that is explicit from the use of the indices $i$ (and will be identified with the holonomy group of the Calabi-Yau manifold) the parts with $s^1+s^2+s^3=\pm 3$ form singlets, and the parts with $s^1+s^2+s^3=\pm 1$ form triplets.

We can split the 32-component $\lambda$ (which satisfies the 10d Weyl condition) in eight 4-component Weyl spinors of 4d according to the $s^i$. We choose the names of the 4d spinors as follows:
\begin{equation}
 \begin{array}{lclclcl}
   \lambda_{+}^\zp & : & s^1=s^2=s^3=+ 1\,,   & \qquad  & \lambda^{\ip}_{+} & : & s^i=+ 1\,,\quad s^j=- 1\ (j\neq i)\,,   \\[2mm]
\lambda_{-}^\zm & : & s^1=s^2=s^3=- 1\,,   &   & \lambda^\im_{-} & : & s^i=- 1\,,\quad s^j=+ 1\ (j\neq i)\,.
   \end{array}
 \label{nameslambdai}
\end{equation}
In the explicit basis of the $8\times 8$ matrices given in \eqref{ED6}, the (4-spinor) components of the 32-component $\lambda$ can be explicitly identified. Since all the $\tilde{E}_{\tilde a}$ are given as tensor product of three $2\times2$ matrices, we can denote spinor components uniquely by the eight triplets $(a_1 a_2 a_3)$ with $a_n \in \{1,2\}$. Here a 1 in the $n$-th place denotes the first row of the $n$-th $2\times2$ matrix and a 2 refers to the second row. This gives
\begin{eqnarray}
  \left(\lambda^{(111)}= \lambda_-^{\zm},\,\lambda^{(112)}= \lambda_+^{\pthree},\, \lambda^{(121)}=\lambda_+^{\ptwo},\, \lambda^{(122)}=\lambda_-^{\mone},\,\right.\nonumber\\
   \left.\lambda^{(211)}=\lambda_+^{\pone},\, \lambda^{(212)}=\lambda_-^{\mtwo},\, \lambda^{(221)}=\lambda_-^{\mthree},\, \lambda^{(222)}=\lambda_+^{\zp}\right)\,.
 \label{componentslambda}
\end{eqnarray}
Note, that in this notation the action of the charge conjugation matrix $C_{(6)}$ (cf. \eqref{ED6}) is simply given by exchanging 1's and 2's in all three entries.

Now we can easily translate spinor bilinears in 10d to spinor bilinears in 4d. Taking into account the charge conjugation matrices (\ref{C1046}), we have (neglecting total derivatives)
\begin{eqnarray}
  \bar \theta^1\Gamma^\mu\nabla _\mu\theta^1&=&  \bar \lambda_-^\zm\gamma^\mu\nabla _\mu\lambda_+^\zp+
   \bar \lambda_-^{\jm}\gamma^\mu\nabla _\mu\lambda_+^{\ip}\delta_{\ip\jm}\,,\nonumber\\
 \frac{\sqrt{2}}{12}\bar \theta^1\Gamma^{mnp}G_{mnp}\theta^1= \frac{\sqrt{2}}{12}\bar \theta^1\tGamma^{\tilde{a}\tilde{b}\tilde{c}}G_{\tilde{a}\tilde{b}\tilde{c}}\theta^1&= &\bar \lambda_+^\zp\lambda_+^\zp G_{123}+\bar \lambda_-^{\zm}\lambda_-^{\zm}G_{\bar 1\bar 2\bar 3}
  +\left( \bar \lambda_+^{\zp}\lambda_+^{\ip}G_{ij\jb}-\bar\lambda_-^{\zm} \lambda_-^{\im}G_{\ib j \jb} \right)\delta^{j \jb}
\nonumber\\
&&+\ft12\left(\bar \lambda_+^{\ip}\lambda_+^{\jp}\varepsilon_{jk\ell}G_{ i \bar k\bar \ell}+\bar \lambda_-^{\im}\lambda_-^{\jm}\varepsilon_{\jb\bar k\bar \ell}G_{\ib k\ell}\right)\delta^{k\bar k}\delta^{\ell\bar \ell}\,.
 \label{gammaGdecomp}
\end{eqnarray}
We can go to curved indices by replacing $\varepsilon_{ijk}$ by $e_i^ue_j^ve_k^w\Omega_{uvw}$ and $\delta_{i\jb}$ by $\rmi e_i^ue_{\jb}^vJ_{u\bar v}$. This leads e.g. to
\begin{eqnarray}
  \frac{\sqrt{2}}{12}\bar \theta^1\Gamma^{mnp}G_{mnp}\theta^1&= &\bar \lambda_+^\zp\lambda_+^\zp \frac{1}{3!} \bar \Omega^{uvw} G_{uvw}+\bar \lambda_-^{\zm}\lambda_-^{\zm}\frac{1}{3!} \Omega^{\bar u\bar v\bar w} G_{\bar u \bar v \bar w}\nonumber\\
   &&
  +\rmi\bar \lambda_+^{\zp}\lambda_+^{\ip}e_i^uG_{uv\bar w}J^{v\bar w}-\rmi\bar\lambda_-^{\zm} \lambda_-^{\im}e_{\ib}^{\bar u}G_{\bar u \bar v w}J^{\bar vw}
\nonumber\\
&&+\ft12\bar \lambda_+^{\ip}\lambda_+^{\jp}e_i^{w}e_j^t\Omega_{uvw}g^{u\bar u}g^{v\bar v}G_{ t \bar u\bar v}
+\ft12\bar \lambda_-^{\im}\lambda_-^{\jm}e_{\im}^{\bar p}e_{\jm}^{\bar w}\bar \Omega_{\bar u\bar v\bar w}g^{u\bar u}g^{v\bar v }G_{\bar p uv}\,.
 \label{gammaGdecompcovariant}
\end{eqnarray}

In the complex coordinates, the Levi-Civita symbol is
\begin{equation}
  \varepsilon^{123\bar 1 \bar 2\bar 3}=\rmi \,,\qquad \varepsilon_{123\bar 1 \bar 2\bar 3}=-\rmi\,.
 \label{LeviCivitacomplex}
\end{equation}
This implies for arbitrary 3-forms $F$ in a flat basis
\begin{eqnarray}
 i\neq j\neq k &:& (*_6 F)_{ijk}= -\rmi F_{ijk}\,,\qquad (*_6 F)_{\ib\jb\bar k}= \rmi F_{\ib\jb\bar k}\,,\nonumber\\
 i\neq j\neq k &:& (*_6 F)_{ij\bar k}= \rmi F_{ij\bar k}\,,\qquad (*_6 F)_{i\bar j\bar k}= -\rmi F_{\ib\jb\bar k}\,,\nonumber\\
 &&\delta^{i\ib}(*_6 F)_{i\ib j}= -\rmi \delta^{i\ib}F_{i\ib j}\,,\qquad \delta^{i\ib}(*_6 F)_{i\ib \jb}= \rmi \delta^{i\ib}F_{i\ib \jb}\,.
 \label{starholom}
\end{eqnarray}
It also allows us to simply express the imaginary self-dual part $G^{\rm ISD}$ of a flux $G$ in terms of the flux $G$ itself. In particular denoting by $(p,q)$ the part of $G$ with $p$ holomorphic and $q$ anti-holomorphic indices (and opposite for $\bar G$) we find:
\begin{eqnarray}
 (3,0) & : & G^{\rm ISD}_{uvw}=\bar G^{\rm ISD}_{\bar u \bar v \bar w} =0\,,\nonumber\\
 (2,1) & : & G^{\rm ISD}_{uv\bar w}=G_{uv\bar w}\,,\qquad \bar G^{\rm ISD}_{u\bar v\bar w}=\bar G_{u\bar v\bar w}\,,\qquad \mbox{primitive}\,,\nonumber\\
 (2,1) & : & G^{\rm ISD}_{uv\bar w}J^{v\bar w}=\bar G^{\rm ISD}_{u\bar v\bar w}J^{u\bar v}=0\,,\nonumber\\
 (1,2) & : & G^{\rm ISD}_{u\bar v\bar w}=\bar G^{\rm ISD}_{uv\bar w}=0\,,\qquad \mbox{primitive}\,,\nonumber\\
 (1,2) & : & G^{\rm ISD}_{u\bar v\bar w} J^{u\bar v}=G_{u\bar v\bar w} J^{u\bar v} \,, \qquad \bar G^{\rm ISD}_{uv\bar w} J^{v\bar w}=\bar G_{u v\bar w} J^{v\bar w}\,,\nonumber\\
 (0,3) & : & G^{\rm ISD}_{\bar u\bar v\bar w}=G_{\bar u \bar v \bar w}\,,\qquad \bar G^{\rm ISD}_{uvw}=\bar G_{uvw}\,.
 \label{ISDcomponents}
\end{eqnarray}

Therefore, the action (\ref{eq:antiD3action2}) becomes in 4 dimensions:
\begin{eqnarray}
{\cal L}_f^{\overline {\rm D3}} &=& T_3 e^{4 A_0} \, \left[
 e^{-\phi} \bar \lambda_-^\zm\gamma^\mu\nabla _\mu\lambda_+^\zp + e^{-\phi}\bar \lambda_-^{\jm}\gamma^\mu\nabla _\mu\lambda_+^{\ip}\delta_{\ip\jm}\right.\label{eq:antiD3actiond4}
\\
 && \left.-\frac{1}{\sqrt{2}}\frac{1}{3!}\rmi\bar \lambda_-^{\zm}\lambda_-^{\zm} \Omega^{\bar u\bar v\bar w} G_{\bar u \bar v \bar w}
  -\frac{1}{\sqrt{2}}\bar \lambda_+^{\zp}\lambda_+^{\ip}e_i^u\bar G_{uv\bar w}J^{v\bar w}-\frac{1}{2\sqrt{2}}\rmi\bar \lambda_-^{\im}\lambda_-^{\jm}e_{\im}^{\bar p}e_{\jm}^{\bar w}\bar \Omega_{\bar u\bar v\bar w}g^{u\bar u}g^{v\bar v }G_{\bar p uv}+\hc\right]\,.\nonumber
\end{eqnarray}

\section{A toroidal example}\label{app:example}

One of the simplest GKP examples is the compactification of type IIB string theory on an orientifold of $T^6/\mathbb{Z}_2 \times \mathbb{Z}_2$, where the $\mathbb{Z}_2 \times \mathbb{Z}_2$ action allows the $T^6$ to factorize into three $T^2$'s. If we furthermore mod out by a $\mathbb{Z}_3$ symmetry that permutes these three $T^2$'s then we are left with three closed string moduli: the complexified volume modulus $\rho$ that controls the volume of the $T^2$'s and involves the integral over the RR form $C_4$, the complex structure modulus $U$ that controls the complex structure of the three identical $T^2$'s and the axio-dilaton $\tau = C_0 +\rmi \, e^{-\phi}$.

Explicitly, on $T^6/\mathbb{Z}_2 \times \mathbb{Z}_2$ we can introduce three complex coordinates $z^u = \frac{1}{\sqrt{2}} (x^{u+3} + U x^{u+6})$, $u=1,2,3$ that are periodic $z^u \sim z^u+1\sim z^u+U$, where the complex $U$ is the complex structure modulus. The $\mathbb{Z}_2 \times \mathbb{Z}_2$ orbifold acts as
\be
\mathbb{Z}_2^{(1)}: (z^1,z^2,z^3) \rightarrow (-z^1,-z^2,z^3)\,, \qquad \mathbb{Z}_2^{(2)}: (z^1,z^2,z^3) \rightarrow (z^1,-z^2,-z^3)\,.
\ee
We furthermore do an orientifold projection and mod out by $\Omega (-1)^{F_L} \mathcal{I}$, where $\Omega$ denotes the string worldsheet parity operator, $F_L$ is the left-moving fermion number and the spacetime involution $\mathcal{I}$ acts on the coordinates as
\be
\mathcal{I}: (z^1,z^2,z^3) \rightarrow (-z^1,-z^2,-z^3)\,.
\ee
There are 64 fixed points for which each of the $z^u$ is either $0, 1/2, U/2$ or $(1+U)/2$. At each of these fixed points sits an O3 plane. We can place a single $\overline {\rm  D3}$ brane on one (or more) of these O3-planes and explicitly calculate its action, which is what we do below.

The total negative D3 brane charge induced by the 64 O3 planes and the $\overline {\rm  D3}$ brane needs to be canceled by turning on $G_{(3)}$ flux of ISD type (recall that ISD flux carries D3 brane charge while IASD flux carries $\overline {\rm  D3}$ brane charge). This ISD flux generically gives a mass to the axio-dilaton modulus $\tau$ and the complex structure modulus $U$. In order to stabilize the volume we need further non-perturbative corrections like a gaugino condensate on D7 branes or Euclidean D3 branes \cite{Kachru:2003aw}. The interplay between the $\overline {\rm  D3}$ brane and the non-perturbative effects can then stabilize all moduli in a dS vacuum \cite{Kachru:2003aw,Kallosh:2004yh}. \footnote{It was recently shown that one can obtain fully analytic dS vacua without the $\overline {\rm  D3}$ brane, if one turns on fluxes that break supersymmetry spontaneously \cite{Kallosh:2014oja}.} We will not explicitly consider these non-perturbative effects and restrict ourselves to a $\overline {\rm  D3}$ brane in the Minkowski solutions with flat volume modulus discovered by %GKP \cite{Giddings:2001yu}.
 \cite{Dasgupta:1999ss,Giddings:2001yu}.

We work for simplicity in the large volume limit in which we can neglect the warp factor. The reason is that the identification of the universal \K \ modulus $\rho$ in warped compactifications is highly non trivial  (see \cite{Frey:2008xw}). The internal unwarped metric and the \K \ form take the simple form
\be
ds^2 = \frac{\text{Im}(\rho)^{\frac12}}{\text{Im}(U)}\sum_{u=1}^3 dz^u d\bar{z}^{\bar u}\,, \qquad J = \rmi \frac{\text{Im}(\rho)^{\frac12}}{\text{Im}(U)} \sum_{u=1}^3 dz^u \wedge d\bar{z}^{\bar u} \,,
\ee
and satisfy $\int d^6 x \sqrt{g} = \frac{\rmi}{3!} \int J \wedge J \wedge  J = \text{Im}(\rho)^{\frac32}$. We normalize the holomorphic $(3,0)$ $\Omega$ such that $\Omega \wedge \bar{\Omega} = -\frac{\rmi}{3!}  J \wedge J \wedge J$, i.e. we take
\be
\Omega = \frac{\text{Im}(\rho)^{\frac34}}{\text{Im}(U)^{\frac32}}\ dz^1 \wedge dz^2 \wedge dz^3\,.
\ee
One can then check (for example by explicit calculation) that the (0,3) $G_{(3)}$ flux and the primitive (2,1) $G_{(3)}$ flux are ISD (cf. \eqref{eq:ISD}). (The (3,0) and the primitive (1,2) parts are IASD while the non-primitive (2,1) and (1,2) fluxes are a combination of ISD and IASD fluxes.)

We now turn on the following ISD $G_{(3)}$ flux
\be
G_{(3)} = n_{(2,1)} \left( dz^1 \wedge dz^2 \wedge d\bar{z}^{\bar 3} + dz^1 \wedge d\bar{z}^{\bar 2} \wedge dz^3 + d\bar{z}^{\bar 1} \wedge dz^2 \wedge dz^3 \right) + n_{(0,3)} d\bar{z}^{\bar{1}} \wedge d\bar{z}^{\bar{2}} \wedge d\bar{z}^{\bar{3}} \,.
\ee
This flux preserves the linearly realized $\mathcal{N}=1$ supersymmetry, iff $n_{(0,3)}=0$ and otherwise breaks all supersymmetry spontaneously. Note, that the Bianchi identity for $F_5$, i.e. the D3 brane charge cancelation, and the flux quantization put constraints on the real parameters $n_{(2,1)}$ and $n_{(0,3)}$ that are however irrelevant for our analysis.

Using the methods of Appendix~\ref{app:d10d4}, we find that the $\overline {\rm D3}$ brane action is given by
\bea
{\cal L}_f^{\overline {\rm D3}} &=2\, T_3\, e^{-\phi} \, \Bigg[& \bar \lambda_-^\zm\gamma^\mu\nabla _\mu\lambda_+^\zp+
  \bar \lambda_-^{\jm}\gamma^\mu\nabla _\mu\lambda_+^{\ip}\delta_{\ip\jm} \\
&&+ \frac{\rmi\, e^{\phi}\ \text{Im}(U)^{\frac32}}{2 \sqrt{2}\ \text{Im}(\rho)^{\frac34}} \left( n_{(0,3)} \left(\bar{\lambda}_+^{0} \lambda_+^{0} -\bar{\lambda}_-^{\bar 0} \lambda_-^{\bar{0}}\right) + n_{(2,1)}  \left(\sum_i \bar{\lambda}_+^{i} \lambda_+^{i} - \sum_{\ib} \bar{\lambda}_-^{\ib} \lambda_-^{\ib}\right) \right) \Bigg]\cr
 &=2\, T_3\, e^{-\phi} \, \Bigg[&\bar \lambda_-^\zm\gamma^\mu\nabla _\mu\lambda_+^\zp+
  \bar \lambda_-^{\jm}\gamma^\mu\nabla _\mu\lambda_+^{\ip}\delta_{\ip\jm} \cr
 &&+ \frac{\sqrt{2}}{24}\,\rmi \,e^\phi \,\bar \Omega^{uvw}\bar G^{{\rm ISD}}_{ u v  w}\ \bar{\lambda}_+^0 \lambda_+^0 - \frac{\sqrt{2}}{24}\,\rmi \,e^\phi \, \Omega^{\bar u\bar v\bar w} G^{{\rm ISD}}_{\bar u\bar v\bar  w}\  \bar{\lambda}_-^{\bar 0} \lambda_-^{\bar 0} \cr
 &&+ \frac{\sqrt{2}}{8}\rmi e^\phi e_i^{w}e_j^t\,\Omega_{uvw}g^{u\bar u}g^{v\bar v}\bar G^{{\rm ISD}}_{t \bar u\bar v}\ \bar{\lambda}_+^i \lambda_+^j -\frac{\sqrt{2}}{8}\rmi e^\phi e_{\ib}^{\bar w} e_{\jb}^{\bar t}\, \bar{\Omega}_{\bar u\bar v\bar w} g^{u\bar u}g^{v\bar v} G^{{\rm ISD}}_{\bar t u v}\ \bar{\lambda}_-^{\ib} \lambda_-^{\jb}  \Bigg]\,.\nonumber
\eea
One can also explicitly check that turning on non-primitive $G_{(3)}$ flux leads to a mixing between $\lambda^0$ and $\lambda^i$, however, such a flux is forbidden by the background equations of motion \cite{Giddings:2001yu}. Thus, we conclude that this example is in complete agreement with the more general results in \eqref{eq:antiD3action4d}-\eqref{eq:mtrip} and provides an explicit example that confirms the results in the main part of the paper.

\section{A note on the no-go theorem for dS space} \label{ss:nogo}

In this paper we studied the bosonic and the fermionic action for a $\overline {\rm D3}$ brane in the presence of fluxes as well as O3 planes. We expect a positive cosmological solution in the four space-time direction and therefore we should discuss how the no-go condition of Gibbons-Maldacena-N\'{u}\~{n}ez \cite{GMN}, recently updated in \cite{Dasgupta:2014pma}, is averted.

Our starting point will be a metric ansatz, originally due to \cite{Dasgupta:1999ss}, specified by a warp factor $e^{2A}$, see (\ref{metricGKP}), with a compact six-dimensional internal manifold, and a general 4d metric $g_{\mu \nu }$ rather than $\eta _{\mu \nu }$. Imposing certain consistency conditions on the background equations of motion we can derive the following constraint equation in the presence of the three-form and the five-form fluxes:
\begin{eqnarray}\label{conds}
 &&~ {V}_6 {R}_4+\int d^6x\;\sqrt{{\tilde{g}_6}}~ {\cal I}_6
 + {\kappa_{10}^2\over 2}\int d^6x\; e^{2A}\sqrt{{\tilde{g}_6}}\Big\{\left[T^k_{k ({\rm O3})}-T^\mu_{\mu ({\rm O3})} \right]\nonumber\\
 && ~~~~~+ \left[T^k_{k ({\rm D3/\overline{\rm D3}})}-T^\mu_{\mu ({\rm D3/\overline{\rm D3}})} \right]\Big\} - \int d^6x\;\sqrt{{\tilde{g}_6}}\, \square \,e^{4A} = 0\,,
\end{eqnarray}
where $R_4 = g^{\mu\nu} R_{\mu\nu}$ is the four-dimensional curvature that would be positive, negative or zero depending on whether the four-dimensional space-time is de-Sitter, anti de-Sitter or Minkowski respectively. The coordinates $x^k$, ($k = 4, 5, ...9$) denote the internal coordinates, and $x^\mu$, ($\mu = 0, 1, 2, 3$) denote the spacetime coordinates. $\square$ is calculated with the internal unwarped metric, i.e. $\sqrt{{\tilde{g}_6}}\, \square=\partial_k \sqrt{{\tilde{g}_6}}\tilde{g}^{k\ell }\partial_\ell$.
The other variables appearing in \eqref{conds} are defined in the following way:
$T^k_{k({\rm O3})}$ denotes the trace of the stress-energy tensor of a O3 plane
localized at some point in the internal space, $T^k_{k({\rm D3}/\overline {\rm D3})}$ denotes the trace of the stress-energy tensor of a D3 or $\overline {\rm D3}$ brane in the internal space; and both ${\cal I}_6$ and $V_6$ are positive definite quantities expressed in terms of type IIB fluxes in \cite{Dasgupta:2014pma} as:
\bea\label{v6I6}
&& {V}_6 \equiv ~ \int d^6x \sqrt{\tilde{g}_6} > 0\,, \nonumber\\
&& {\cal I}_6 \equiv ~ \frac{e^{2A}G_{(3)} \cdot \bar{G}_{(3)}}{12\; {\rm
    Im}\tau} -\frac{e^{2A}F_{(5)} \cdot F_{(5)}}{4 \cdot 4!}+e^{-6A}\partial_m e^{4A}\partial^m e^{4A}\ge 0\,.
\eea
$G_{(3)} \cdot \bar{G}_{(3)}$ is calculated with the full metric, i.e. $G_{(3)} \cdot \bar{G}_{(3)}=e^{6A}G_{k\ell m}\tilde{g}^{kk'}\tilde{g}^{\ell \ell '}\tilde{g}^{mm'}\bar G_{k'\ell 'm'}$. Here $F_{(5)} \cdot F_{(5)}=F_{(5)\mu k\ell mn }  F_{(5)}^{\mu k\ell mn}$, where indices are raised again with the full metric.
Our aim now is to see under what condition we can get a positive curvature solution $R_4 > 0$ using the constraint equation \eqref{conds}.
Looking at \eqref{conds}, we see that, in the presence of D3 branes, $\overline {\rm D3}$ branes, O3 planes and fluxes, there are three possible cases that could potentially arise here.

\noindent $\bullet$ {The warp factor $e^{2A}$ is smooth and the internal six-dimensional manifold is compact}.

\noindent $\bullet$ {The internal six-dimensional manifold is compact, but the warp factor is not smooth}.

\noindent $\bullet$ {The internal six-dimensional manifold is non-compact and the warp-factor may or may not be smooth}.

\noindent For all these cases, as discussed in \cite{Dasgupta:2014pma}, the conclusions are similar. It is impossible to satisfy \eqref{conds} without invoking perturbative and/or non-perturbative quantum corrections. These corrections, in the form of instantons and multi-instantons corrections in type IIB, are somehow necessary to support a positive curvature solution in four spacetime dimensions. This is of course
the KKLT paradigm \cite{Kachru:2003aw} where these corrections are automatically generated once we stabilize the K\"{a}hler and complex structure moduli of the internal six-dimensional manifold, including the dilaton and the moduli of the seven-branes.
The compact six-dimensional manifold is equipped with the right amount of fluxes etc. that is necessary for global charge cancelation. All the stabilizing effects, for example quantum corrections etc., can be thought of as being added to generate a four-dimensional metric with a cosmological constant $\Lambda$ and an internal space with a time-independent warp factor $e^{2A}$. This way the size of the internal space will be time-independent, implying the time-independency of the four-dimensional Newton constant.

The constraint on the cosmological constant $\Lambda$ and the warp factor $e^{2A}$ in the presence of type IIB branes, planes, fluxes and quantum corrections can be easily studied from an M-theory set-up where the analysis is relatively accessible\footnote{In M-theory all the type IIB fluxes can be neatly packaged in the 4-form ${\cal G}$-flux. The type IIB branes map to either geometry or M2 and M5 branes.
Additionally the instantons and other non-perturbative corrections in type IIB map to certain generalized curvature corrections in M-theory. In this language it is therefore somewhat easier to deal with M-theory than type IIB. Of course in either case, the physics remains unchanged.}.
In M-theory the internal space is eight-dimensional which can be thought of as a $T^2$ fibration over our six-dimensional space with an unwarped metric $g_{kl}$ discussed earlier\footnote{Recall that in the limit
when the size of the $T^2$ vanishes, M-theory reduces to type IIB theory. This way a $T^2$-fibered eight dimensional manifold in M-theory, with a six-dimensional base specified by the metric $g_{kl}$,
reduces to type IIB on the six-dimensional base.}.

The constraint in M-theory now, after adding all the necessary ingredients, is a variant of \eqref{conds} and is given by \cite{Dasgupta:2014pma}:
\begin{eqnarray}\label{conokl}
&& {1\over 12}  \int d^8 y \sqrt{\tilde{g}_8}~{\cal G}_{(4)} \cdot{\cal G}_{(4)}  + 12 \Lambda \int d^8 y \sqrt{\tilde{g}_8}~ e^{-8A}  + 2 \kappa^2 T_2 (n_3 + {\overline{n}}_3)\nonumber\\
&& ~~~~~~~~~~ + \left[\langle {\cal T}^k_k\rangle_q - \langle {\cal T}^\mu_\mu\rangle_q\right] + \int d^8 y \sqrt{\tilde{g}_8}~\square\, e^{-4A} ~ = ~ 0,
\end{eqnarray}
where the Roman letters now span the internal eight-dimensional space and the Greek letters span the $2+1$ dimensional spacetime. The other notations are as follows:
$\tilde{g}_{8}$ above is the uplift of the six-dimensional internal metric to eleven-dimensions, and ${\cal G}_{(4)}$ is the uplift of the type IIB three and five-form fluxes to eleven-dimensions such that ${\cal G}_{(4)} \cdot{\cal G}_{(4)} = G_{mnpa}G^{mnpa}$, where
($m, n, p$) span the coordinates of the six-dimensional base and ($a, b$) span the coordinates of the $T^2$ fiber. The raising and lowering of indices are performed
using unwarped M-theory metric.
The branes and anti-branes are denoted by $n_3$ and $\overline{n}_3$ respectively, and the O-planes become smooth spaces in M-theory\footnote{For example the O3 and O5 planes become orbifold points that can be resolved using certain ``twisted-sector'' states in M-theory \cite{DM, witten95}.
The O7-plane either becomes a smooth Atiyah-Hitchin space or a Horava-Witten wall. The O9-plane is more complicated but this can also be mapped to the Horava-Witten wall. \label{opla}}.
The energy-momentum tensor from the quantum corrections i.e the curvature corrections (see \cite{Dasgupta:2014pma} for details) are denoted by $\langle {\cal T}^k_k\rangle_q$.

Since the warp factor $e^{2A}$ is smooth in M-theory, for a compact eight-dimensional manifold, we expect the integral of $\square\, e^{-4A}$ to vanish. In this case it is easy to see that the $\Lambda > 0$ condition can be achieved if and only if:
\begin{equation}\label{fincond}
\langle {\cal T}^\mu_\mu\rangle_q ~ > ~  \langle {\cal T}^k_k\rangle_q,
\end{equation}
which is the generalization of the classical condition first found by \cite{GMN}. The analysis that we perform here, in the presence of the above-mentioned corrections, would precisely achieve that, allowing a positive
curvature solution to exist in our set-up.

\end{document}